\documentclass[epj,nopacs,final]{preprint}
\usepackage{epsfig}
\gdef\Feynmanlength{\setlength{\unitlength}{0.01pt}}  

\global\newcount\LINETYPE                     
\global\newcount\LINEDIRECTION
\global\newcount\LINECONFIGURATION
\newcommand{\LTYPE}{\LINETYPE}
\newcommand{\LDIR}{\LINEDIRECTION}

\global\LINETYPE=1  \global\LINEDIRECTION=0  \global\LINECONFIGURATION=0
\global\newcount\fermion    \fermion=1
\global\newcount\scalar     \scalar=2
\global\newcount\photon     \photon=3
\global\newcount\gluon      \gluon=4
\global\newcount\especial   \especial=5
    \gdef\E{2}   
    \gdef\W{6}   
\global\newcount\REG            \global\REG=0
\global\newcount\FLIPPED        \global\FLIPPED=1
\global\newcount\CURLY          \global\CURLY=2
\global\newcount\FLIPPEDCURLY   \global\FLIPPEDCURLY=3
\global\newcount\FLAT           \global\FLAT=4
\global\newcount\FLIPPEDFLAT    \global\FLIPPEDFLAT=5
\global\newcount\CENTRAL        \global\CENTRAL=6
\global\newcount\FLIPPEDCENTRAL \global\FLIPPEDCENTRAL=7
             
\global\newcount\SQUASHEDGLUON  \global\SQUASHEDGLUON=8

\newcount\adjx \adjx=0
\newcount\adjy \adjy=0
\global\newdimen\BIGPHOTONS     \BIGPHOTONS=0pt  
\global\newdimen\THICKPHOTONS     \THICKPHOTONS=0pt  
\global\newdimen\THICKPHOTONSWITCH    \THICKPHOTONSWITCH=0pt
\gdef\THICKPHOTONTEST{
\THICKPHOTONSWITCH=0pt
\ifdim\THICKPHOTONS=0pt \relax
  \else \ifnum\LTYPE=3
           \ifnum\LDIR=2 \THICKPHOTONSWITCH=1pt \fi 
           \ifnum\LDIR=6 \THICKPHOTONSWITCH=1pt \fi 
        \fi
\fi
}  

\global\newcount\phantomswitch   \global\phantomswitch=0
\global\newcount\stemlength   \global\stemlength=275   
\global\newcount\absstemlength        
\global\newcount\stemlengthx          
\global\newcount\stemlengthy          
\newdimen\FRONTSTEM  \FRONTSTEM=0pt   
\newdimen\BACKSTEM   \BACKSTEM=0pt    
\newdimen\EITHERSTEM \EITHERSTEM=0pt  
\global\newcount\arrowlength                
\global\newdimen\ATTIP   \global\ATTIP=0pt  
\global\newdimen\ATBASE  \global\ATBASE=1pt 
\global\newcount\unitboxnumber  
\global\newcount\unitboxnumberpo  
\global\newcount\particlelengthx  
\gdef\plengthx{\particlelengthx}
\global\newcount\particlelengthy  
\gdef\plengthy{\particlelengthy}  
\global\newcount\boxlengthx  
\global\newcount\boxlengthy  
\global\newcount\particleadjustx  
\global\newcount\particleadjusty  
\global\newcount\particlelength   
\global\newcount\particlefrontx
\gdef\pfrontx{\particlefrontx}
\global\newcount\PFRONTx
\global\newcount\particlefronty
\gdef\pfronty{\particlefronty}
\global\newcount\PFRONTy
\global\newcount\particlebackx
\gdef\pbackx{\particlebackx}
\global\newcount\particlebacky
\gdef\pbacky{\particlebacky}
\global\newcount\particlemidx
\gdef\pmidx{\particlemidx}
\global\newcount\particlemidy
\gdef\pmidy{\particlemidy}
\global\newcount\seglength  \global\newcount\gaplength
\global\gaplength=850  
\global\seglength=1416  
\global\newcount\Xone    \global\newcount\Yone    
\global\newcount\Xtwo    \global\newcount\Ytwo    
\global\newcount\Xthree  \global\newcount\Ythree  
\global\newcount\Xfour   \global\newcount\Yfour   
\global\newcount\Xfive   \global\newcount\Yfive   
\global\newcount\Xsix    \global\newcount\Ysix    
\global\newcount\Xseven  \global\newcount\Yseven  
\global\newcount\Xeight  \global\newcount\Yeight  
\newsavebox{\lastline}  
\global\newcount\numlineparts   
\global\newcount\upperlineadjx  \upperlineadjx=0  
\global\newcount\upperlineadjy  \upperlineadjy=0  
\global\newcount\lowerlineadjx  \lowerlineadjx=0  
\global\newcount\lowerlineadjy  \lowerlineadjy=0  
\global\newcount\thirdlineadjx  \thirdlineadjx=0  
\global\newcount\thirdlineadjy  \thirdlineadjy=0  
\global\newcount\fourthlineadjx \fourthlineadjx=0  
\global\newcount\fourthlineadjy \fourthlineadjy=0  
\global\newcount\unitboxwidth   \unitboxwidth=1000
\global\newcount\unitboxheight  \unitboxheight=0  
\global\newcount\numupperunits  \numupperunits=8  
\global\newcount\numlowerunits  \numlowerunits=8  
\global\newcount\numthirdunits  \numthirdunits=8  
\global\newcount\numfourthunits \numfourthunits=8  
\global\newcount\fermioncount   \global\fermioncount=0    
\global\newcount\scalarcount    \global\scalarcount=0    
\global\newcount\photoncount    \global\photoncount=0    
\global\newcount\gluoncount     \global\gluoncount=0    
\global\newcount\especialcount  \global\especialcount=0    
\global\newcount\vertexcount    \global\vertexcount=-1
\global\newcount\XDIR
\global\newcount\YDIR
\gdef\SETDIR{  
\ifcase\LDIR 
     \global\XDIR=0  \global\YDIR=1   
\or  \global\XDIR=1  \global\YDIR=1   
\or  \global\XDIR=1  \global\YDIR=0   
\or  \global\XDIR=1  \global\YDIR=-1  
\or  \global\XDIR=0  \global\YDIR=-1  
\or  \global\XDIR=-1 \global\YDIR=-1  
\or  \global\XDIR=-1 \global\YDIR=0   
\or  \global\XDIR=-1 \global\YDIR=1   
\else\DIRECTERROR 
\fi}  
\gdef\moduloeight#1{
\ifnum#1>7 \global\advance #1 by -8 
\relax
\moduloeight#1 
\relax
\else \relax  
\fi}
\gdef\multroothalf#1{\global\multiply #1 by 7071 \global\divide #1 by 10000}
\gdef\negate#1{\global\multiply #1 by -1}

\gdef\slanttest(#1,#2){ 
\ifodd\LDIR
\multiply #1 by 7071  \divide #1 by 10000
\multiply #2 by 7071  \divide #2 by 10000
\fi
}
\gdef\gslanttest(#1,#2){
\ifodd\LDIR
\multroothalf#1
\multroothalf#2
\fi
}
\gdef\setplength{ 
\global\particlelengthx=\unitboxwidth
\global\particlelengthy=\unitboxheight
\global\multiply \particlelengthx by \unitboxnumber
\global\multiply \particlelengthy by \unitboxnumber
\global\advance \particlelengthx by \particleadjustx
\global\advance \particlelengthy by \particleadjusty
}
\gdef\boxlengthdefault{  
\global\boxlengthx=\plengthx
\global\boxlengthy=\plengthy
\ifnum\plengthx<0 \global\multiply\boxlengthx by -1 \fi
\ifnum\plengthy<0 \global\multiply\boxlengthy by -1 \fi
}
\gdef\rearcoords{  
\global\particlebacky=\particlefronty 
\global\particlebackx=\particlefrontx 
\global\advance \particlebackx by \particlelengthx
\global\advance \particlebacky by \particlelengthy
}
\gdef\midcoords{  
\global\particlemidy=\particlefronty
\global\particlemidx=\particlefrontx
\global\stemlengthx=\particlelengthx  
\global\stemlengthy=\particlelengthy  
\global\divide\stemlengthx by 2
\global\divide\stemlengthy by 2
\global\advance \particlemidx by \stemlengthx
\global\advance \particlemidy by \stemlengthy
}
\gdef\setcoords(#1,#2,#3)(#4,#5,#6)[#7,#8]{  
\global\upperlineadjx=#1
\global\lowerlineadjx=#2
\global\thirdlineadjx=#3
\global\upperlineadjy=#4
\global\lowerlineadjy=#5
\global\thirdlineadjy=#6
\global\unitboxwidth=#7
\global\unitboxheight=#8
}
\gdef\drawoldpic#1(#2,#3){  
\global\particlefrontx=#2
\global\particlefronty=#3
\rearcoords  
\midcoords
\put(#2,#3){\usebox{#1}}
}
\gdef\drawsavedline`#1' as #2[#3#4](#5,#6)[#7]{
\global\LINETYPE=#2
\global\LINEDIRECTION=#3
\global\LINECONFIGURATION=#4
\global\particlefrontx=#5
\global\particlefronty=#6
\global\unitboxnumber=#7  
\selectcase
\rearcoords
\midcoords
\ifnum\phantomswitch=0 \drawas{#1}\fi
}

\gdef\drawas#1{
\global\savebox{#1}(\boxlengthx,\boxlengthy){
\setlength{\unitlength}{0.01pt}
\begin{picture}(\boxlengthx,\boxlengthy)
\multiput(\upperlineadjx,\upperlineadjy)(\unitboxwidth,\unitboxheight)
{\numupperunits}{\upperunitbox}
\ifnum\numlineparts > 1  
\multiput(\lowerlineadjx,\lowerlineadjy)(\unitboxwidth,\unitboxheight)
{\numlowerunits}{\lowerunitbox}  
\fi
\ifnum\numlineparts > 2  
\multiput(\thirdlineadjx,\thirdlineadjy)(\unitboxwidth,\unitboxheight)
{\numthirdunits}{\thirdunitbox}  
\fi
\ifnum\numlineparts > 3  
\multiput(\fourthlineadjx,\fourthlineadjy)(\unitboxwidth,\unitboxheight)
{\numfourthunits}{\lowerunitbox}  
\fi
\end{picture} }
\global\PFRONTx=\pfrontx  \global\PFRONTy=\pfronty   
\SETFRONTSTEM
\THICKPHOTONTEST
\ifdim\THICKPHOTONSWITCH=1pt\global\advance\PFRONTy by 20  \fi
\put(\PFRONTx,\PFRONTy) {\usebox{#1}}   
\ifdim\THICKPHOTONSWITCH=1pt
\global\advance\PFRONTy by -40
\put(\PFRONTx,\PFRONTy) {\usebox{#1}}   
\global\advance \PFRONTy by 20  
\fi  
\SETBACKSTEM
\seglength=1416   \gaplength=850   
}
\gdef\drawandsaveline`#1' as #2[#3#4](#5,#6)[#7]{
\global\newsavebox{#1}
\drawsavedline`#1' as #2[#3#4](#5,#6)[#7]
}

\gdef\drawline#1[#2#3](#4,#5)[#6]{   
\drawsavedline`\lastline' as #1[#2#3](#4,#5)[#6]}

\gdef\TYPEERROR{\message{*** ERROR IN PARTICLE TYPE SELECTION ***}
\message{+++ Try with line type \fermion,\scalar,\photon,\gluon 
(see manual) +++}\SETERR}
\gdef\DIRECTERROR{\SETERR\message{*** ERROR IN PARTICLE DIRECTION SELECTION ***}
\message{+++ Try again with direction N, NE, E, SE  etc. or see manual +++}}
\gdef\UNIMPERROR{\message{*** ERROR IN PARTICLE OPTIONS SELECTION ***}
\message{
+++ The requested options combination has not yet been implemented +++}\SETERR}
\gdef\SETERR{\gdef\upperunitbox{{\tiny Error}}  
\gdef\lowerunitbox{\relax}
\gdef\thirdunitbox{\relax}
}
\gdef\neglengthcheck{\ifnum\unitboxnumber < 1 
\message{   *** ERROR:  PARTICLE OF NEGATIVE OR ZERO LENGTH REQUESTED. ***   }
\message{   ***         TAKING ABSOLUTE VALUE. ***   }\negate\unitboxnumber \fi}
\gdef\selectcase{  
\neglengthcheck   
\SETDIR  
\ifcase\LINETYPE
\TYPEERROR  
\or \selectfermion  
\or \selectscalar   
\or \selectphoton   
\or \selectgluon    
\or \selectespecial 
\else \TYPEERROR \fi  }
\gdef\selectfermion{
\ifnum\fermioncount=0 \input fermionsetup \fi   
\global\advance\fermioncount by 1  
\ALLfermion   
}
\gdef\selectscalar{
\ifnum\scalarcount=0 \input scalarsetup \fi   
\global\advance\scalarcount by 1  
\ALLscalar
}
\gdef\selectphoton{   
\ifnum\photoncount=0 \input photonsetup  \fi   
\selectphoton
}
\gdef\selectgluon{   
\ifnum\gluoncount=0 \input gluonsetup  \fi
\selectgluon
}
\gdef\selectespecial{\UNIMPERROR}
\gdef\checkvertex{ 
\ifnum\vertexcount=-1   \input vertex  \fi}
\gdef\drawvertex#1[#2#3](#4,#5)[#6]{\checkvertex\drawvertex#1[#2#3](#4,#5)[#6]}
\gdef\vertexcap#1{\checkvertex\vertexcap#1}
\gdef\vertexcaps{\checkvertex\vertexcaps}
\gdef\vertexlink#1{\checkvertex\vertexlink#1}
\gdef\vertexlinks{\checkvertex\vertexlinks}
\gdef\stemvertex#1{\checkvertex\stemvertex#1}
\gdef\stemvertices{\checkvertex\stemvertices}
\gdef\flipvertex{\checkvertex\flipvertex}
\global\arrowlength=349  
\gdef\drawarrow[#1#2](#3,#4){
\global\LDIR=#1
\SETDIR
\global\boxlengthx=#3  
\global\boxlengthy=#4  
\ifdim#2=1pt  
\adjx=\arrowlength      \adjy=\arrowlength
\multiply\adjx by \XDIR \multiply\adjy by \YDIR  
\slanttest(\adjx,\adjy)
\global\advance\boxlengthx by \adjx    \global\advance\boxlengthy by \adjy
\fi
\ifnum\phantomswitch=0\put(\boxlengthx,\boxlengthy){\vector(\XDIR,\YDIR){0}}\fi
}  
\gdef\SETFRONTSTEM{
\EITHERSTEM=\FRONTSTEM   \advance\EITHERSTEM by \BACKSTEM
\ifdim\EITHERSTEM>0pt
\global\stemlengthx=\stemlength   \global\stemlengthy=\stemlength   
\global\absstemlength=\stemlength   
\SETDIR
\gslanttest(\stemlengthx,\stemlengthy)
\gslanttest(\absstemlength,\REG)  
\ifnum\XDIR=0 \stemlengthx=0 \fi
\ifnum\YDIR=0 \stemlengthy=0 \fi
\global\multiply\stemlengthx by \XDIR
\global\multiply\stemlengthy by \YDIR
\ifdim\FRONTSTEM=1pt 
\ifnum\phantomswitch=0
          \put(\pfrontx,\pfronty){\line(\XDIR,\YDIR){\absstemlength}}\fi
\global\advance\plengthx by \stemlengthx
\global\advance\plengthy by \stemlengthy
\global\advance\PFRONTx by \stemlengthx   
\global\advance\PFRONTy by \stemlengthy
\global\advance\pmidx by \stemlengthx
\global\advance\pmidy by \stemlengthy
\global\advance\pbackx by \stemlengthx
\global\advance\pbacky by \stemlengthy
\ifnum\LTYPE=3
\global\photonfrontx=\PFRONTx  \global\photonfronty=\PFRONTy
\global\photonbackx=\pbackx    \global\photonbacky=\pbacky
\fi  
\ifnum\LTYPE=4
\global\gluonfrontx=\PFRONTx  \global\gluonfronty=\PFRONTy
\global\gluonbackx=\pbackx    \global\gluonbacky=\pbacky
\fi  
\fi  
\fi  
}    
\gdef\SETBACKSTEM{
\ifdim\BACKSTEM=1pt 
\ifnum\phantomswitch=0
       \put(\pbackx,\pbacky){\line(\XDIR,\YDIR){\absstemlength}}\fi
\global\advance\plengthx by \stemlengthx
\global\advance\plengthy by \stemlengthy
\global\advance\pbackx by \stemlengthx
\global\advance\pbacky by \stemlengthy
\fi  
\global\stemlength=275  \FRONTSTEM=0pt  \BACKSTEM=0pt 
}    
\gdef\drawloop#1[#2#3](#4,#5){  
\input loops  
\drawloop#1[#2#3](#4,#5)}
\Feynmanlength  

\begin{document}
\preprint {ANL-HEP-PR-98-91}
\title    {Three jet cross sections in photoproduction at HERA}
\author   {Michael Klasen}
\institute{High Energy Physics Division,
           Argonne National Laboratory,
           Argonne, Illinois 60439,
           U.S.A.,
           \email{klasen@hep.anl.gov}}
\date{\today}
\abstract{
We calculate three jet cross sections in photoproduction using exact matrix
elements for the direct and resolved contributions. Numerical distributions
are presented in a generic, irreducible set of variables that allows to
disentangle the dynamics of partonic QCD subprocesses from each other and
from pure phase space distributions. The results are compared to preliminary
data from the ZEUS collaboration at HERA. It is found that the largest
contribution comes from photon-gluon
fusion in the mass range 36 GeV $< M_{\rm 3-jet} <$ 80 GeV.
The measured leading jet scattering angle distribution is consistent with the
$t$-channel exchange of a massless fermion in $2\rightarrow 2$ scattering,
where the third parton is assumed to arise from soft bremsstrahlung. The data
are inconsistent with pure phase space and Rutherford scattering distributions.
}
\maketitle

\section{Introduction}
\label{sec:1}

Photoproduction of jets has been studied to a great extent at HERA since the
DESY electron-proton collider turned on in 1992 by the H1 \cite{H11992} and
ZEUS \cite{ZEUS1992} collaborations. Until recently, however, the limited
luminosity only allowed for measuring inclusive single jet and dijet cross
sections. In general, the data have shown good agreement with QCD predictions
accurate to next-to-leading order in perturbation theory \cite{Klasen}. They
have confirmed the existence of direct and resolved partonic contributions for
real and slightly off-shell photons and begin to show sensitivity to different
parametrizations of the parton densities in the photon provided that
uncertainties from the underlying hadronic event and the jet definition are
under control \cite{H1Dijet,ZEUSDijet}.

On the other hand, multijet production has been measured some time ago in
purely hadronic collisions at the CERN SPS collider by the UA1 collaboration
\cite{UA1Multijet} and at the Fermilab Tevatron collider by the CDF
\cite{CDFMultijet} and D0 \cite{D0Multijet} collaborations. Its importance
lies not only in testing perturbative QCD to higher orders, but also in the
search for new phenomena. Many analyses for the production of Standard Model
and new particles, e.g. the top quark, the Higgs boson, or supersymmetric
particles, rely on many hard jets in the final state. QCD multijet production
then is a significant background for these searches and has to be well
understood.

At HERA the single jet transverse energy and dijet mass distributions in
photoproduction have shown no deviations from the next-to-leading order QCD
predictions so far. However, higher integrated luminosity has now permitted
ZEUS to produce the first three jet analysis \cite{Brussels}. Even a limited
number of four jet events has been observed. It is therefore interesting to
compare these new data with QCD predictions and look for deviations as
signals of new physics. Since next-to-leading order cross sections for the
photoproduction of three jets are not available, such an analysis is still
restricted to leading order in perturbation theory and will therefore be
subject to large renormalization and factorization scale uncertainties. As a
possible way out, we will concentrate in this paper on normalized
distributions that are largely independent of both higher order corrections
and scale choices as are the shapes of the distributions.
We briefly review the leading order cross section
formalism and the relevant kinematic variables before we present numerical
results for both dynamic QCD and pure kinematic phase space distributions.
We disentangle the different partonic subprocesses that contribute to
photoproduction of three jets and show that the leading jet scattering angle
distribution in three jet production is closely related to two jet production.

The paper is organized as follows: In Sect.~\ref{sec:2}, we briefly review
the leading order $2\rightarrow 3$ parton scattering processes. In
Sect.~\ref{sec:3} we define the relevant phase space variables and the
hadronic three jet cross sections. Our numerical results are presented in
Sect.~\ref{sec:4}, and we summarize our analysis in Sect.~\ref{sec:5}.

\section{Leading order $2\rightarrow 3$ parton scattering}
\label{sec:2}

Our predictions for three jet cross sections in photoproduction are based on
the leading order $2\rightarrow 3$ matrix elements $\overline{|{\cal M}|^2}$
averaged over initial and summed over final spin and color states. The
partonic cross section can then be calculated through
\begin{equation}
 \sigma_{12}^{345} = \frac{1}{2s} \overline{|{\cal M}|^2} (2\pi)^4
 \delta(p_1+p_2-p_3-p_4-p_5)\prod_{i=3}^5 \frac{\mbox{d}^4p_i}{(2\pi)^3}
 \delta(p_i^2).
\end{equation}
We denote the four-momenta of the incoming and outgoing partons by $p_{1,2}$
and $p_{3,4,5}$ and the partonic center-of-mass energy squared
by $s=(p_1+p_2)^2$.

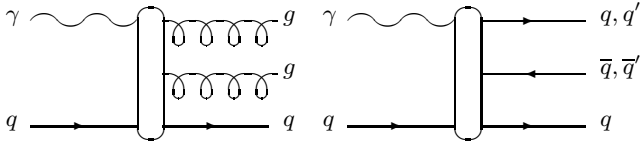
\begin{figure}
\begin{center}
\begin{picture}(23000,6000)
\drawline\photon[\E\REG](1000,5000)[4]
\drawline\fermion[\E\REG](1000,1000)[4000]
\drawarrow[\E\ATTIP](3000,1000)
\put(5500,3000){\oval(1000,5000)}
\drawline\gluon[\E\REG](6000,5000)[4]
\drawline\gluon[\E\REG](6000,3000)[4]
\drawline\fermion[\E\REG](6000,1000)[4000]
\drawarrow[\E\ATTIP](8000,1000)
\put(0,1000){$q$}
\put(0,5000){$\gamma$}
\put(10500,1000){$q$}
\put(10500,5000){$g$}
\put(10500,3000){$g$}
\drawline\photon[\E\REG](13000,5000)[4]
\drawline\fermion[\E\REG](13000,1000)[4000]
\drawarrow[\E\ATTIP](15000,1000)
\put(17500,3000){\oval(1000,5000)}
\drawline\fermion[\E\REG](18000,5000)[4000]
\drawarrow[\E\ATTIP](20000,5000)
\drawline\fermion[\E\REG](18000,3000)[4000]
\drawarrow[\W\ATBASE](20000,3000)
\drawline\fermion[\E\REG](18000,1000)[4000]
\drawarrow[\E\ATTIP](20000,1000)
\put(12000,1000){$q$}
\put(12000,5000){$\gamma$}
\put(22500,1000){$q$}
\put(22500,5000){$q,q'$}
\put(22500,3000){$\overline{q},\overline{q}'$}
\end{picture}
\end{center}
\caption{Leading order $2\rightarrow 3$ Feynman diagrams for direct
         photoproduction.}
\label{fig:1}
\end{figure}

Both direct and resolved processes have to be taken into account. The generic
diagrams for direct $2\rightarrow 3$ scattering, where the photon interacts
directly with a parton in the proton, are displayed in Fig.~\ref{fig:1}.
These diagrams arise from the leading order $2\rightarrow 2$ QCD Compton
scattering process
\begin{equation}
 \overline{|{\cal M}|^2}_{\gamma q \rightarrow q g} =
 16 \pi^2 \alpha\alpha_s 2 C_F \left(-\frac{s}{t}-\frac{t}{s}\right)
\end{equation}
through the emission of an additional gluon (left diagram in Fig.~\ref{fig:1})
or the splitting of the final state gluon into a quark-antiquark pair (right
diagram in Fig.~\ref{fig:1}). $\alpha$ and $\alpha_s$ denote the
electromagnetic and strong coupling constants, $s$, $t$, and $u$ are the usual
Mandelstam variables for $2\rightarrow 2$ scattering, and $C_F=(N_C^2-1)/
(2N_C)$ with $N_C=3$ is the SU(3) color factor. It is worth noting that
both the QCD Compton and the crossed photon-gluon fusion processes proceed
through a massless fermion exchange in the $t$-channel, which leads to a
single $1/t$ pole in the matrix element. Both diagrams in
Fig.~\ref{fig:1} lead to an extra factor of $\alpha_s$, when the matrix
elements are squared, so that the direct $2\rightarrow 3$ cross sections are
of ${\cal O} (\alpha\alpha_s^2)$. The outgoing quark-antiquark pair can be of
the same or different flavor than the incoming quark. The diagrams for
incoming anti-quarks or gluons can be obtained from Fig.~\ref{fig:1} by
crossing a final quark or gluon line with the incoming quark line. All possible
topologies and orders of outgoing particles have to be considered for the full
matrix elements of the processes, although they are not shown here explicitly.
The complete result in $d$ dimensions can be found in \cite{Aurenche} in a very
compact notation. Since there are no singularities present in three jet
production, we can simply set $d=4$.

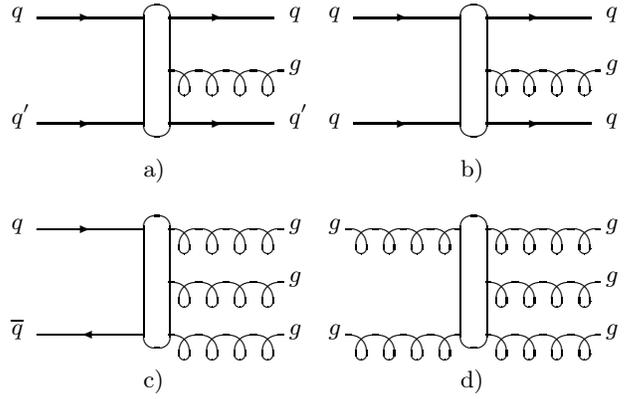
\begin{figure}
\begin{center}
\begin{picture}(23000,15000)
\drawline\fermion[\E\REG](1000,14000)[4000]
\drawarrow[\E\ATTIP](3000,14000)
\drawline\fermion[\E\REG](1000,10000)[4000]
\drawarrow[\E\ATTIP](3000,10000)
\put(5500,12000){\oval(1000,5000)}
\drawline\fermion[\E\REG](6000,14000)[4000]
\drawarrow[\E\ATTIP](8000,14000)
\drawline\fermion[\E\REG](6000,10000)[4000]
\drawarrow[\E\ATTIP](8000,10000)
\put(0,10000){$q'$}
\put(0,14000){$q$}
\put(10500,10000){$q'$}
\put(10500,14000){$q$}
\put(5000,8000){a)}
\drawline\gluon[\E\REG](6000,12000)[4]
\put(10500,12000){$g$}
\drawline\fermion[\E\REG](13000,14000)[4000]
\drawarrow[\E\ATTIP](15000,14000)
\drawline\fermion[\E\REG](13000,10000)[4000]
\drawarrow[\E\ATTIP](15000,10000)
\put(17500,12000){\oval(1000,5000)}
\drawline\fermion[\E\REG](18000,14000)[4000]
\drawarrow[\E\ATTIP](20000,14000)
\drawline\fermion[\E\REG](18000,10000)[4000]
\drawarrow[\E\ATTIP](20000,10000)
\put(12000,10000){$q$}
\put(12000,14000){$q$}
\put(22500,10000){$q$}
\put(22500,14000){$q$}
\put(17000,8000){b)}
\drawline\gluon[\E\REG](18000,12000)[4]
\put(22500,12000){$g$}
\drawline\fermion[\E\REG](1000,6000)[4000]
\drawarrow[\E\ATTIP](3000,6000)
\drawline\fermion[\E\REG](1000,2000)[4000]
\drawarrow[\W\ATBASE](3000,2000)
\put(5500,4000){\oval(1000,5000)}
\drawline\gluon[\E\REG](6000,6000)[4]
\drawline\gluon[\E\REG](6000,2000)[4]
\put(0,2000){$\overline{q}$}
\put(0,6000){$q$}
\put(10500,2000){$g$}
\put(10500,6000){$g$}
\put(5000,0){c)}
\drawline\gluon[\E\REG](6000,4000)[4]
\put(10500,4000){$g$}
\drawline\gluon[\E\REG](12700,6000)[4]
\drawline\gluon[\E\REG](12700,2000)[4]
\put(17500,4000){\oval(1000,5000)}
\drawline\gluon[\E\REG](18000,6000)[4]
\drawline\gluon[\E\REG](18000,2000)[4]
\put(12000,2000){$g$}
\put(12000,6000){$g$}
\put(22500,2000){$g$}
\put(22500,6000){$g$}
\put(17000,0){d)}
\drawline\gluon[\E\REG](18000,4000)[4]
\put(22500,4000){$g$}
\end{picture}
\end{center}
\caption{Leading order $2\rightarrow 3$ Feynman diagrams for resolved
         photoproduction.}
\label{fig:2}
\end{figure}

If the photon resolves into its partonic content before the hard scattering
takes place, the effective hard scattering is of purely partonic nature. The
relevant generic diagrams are displayed in Fig.~\ref{fig:2}.
They are of ${\cal O} (\alpha_s^3)$ and arise from the underlying
$2\rightarrow 2$ parton-parton scattering processes through radiation of
an additional gluon in the final state. Most of the $2\rightarrow 2$
parton-parton processes proceed through the exchange of a massless vector
boson in the $t$-channel leading to a double pole $1/t^2$ in the matrix
elements. For example, the matrix element for the process $qq'\rightarrow
qq'$ is given by
\begin{equation}
 \overline{|{\cal M}|^2}_{qq' \rightarrow qq'} =
 16 \pi^2 \alpha_s^2 \frac{C_F}{N_C} \frac{s^2+u^2}{t^2}.
\end{equation}
Crossing one or two outgoing gluon lines into the initial state in
Fig.~\ref{fig:2} leads to the gluon initiated processes with quarks in
the final state, and crossing an incoming and outgoing quark line
leads to the processes with incoming anti-quarks. Process c) is symmetric
under the interchange of the three gluons and under the interchange of the
quark and anti-quark. Process d) is completely symmetric under the
interchange of any of the five gluons. The complete result for the
$2\rightarrow 3$ parton scattering matrix elements in $d$ dimensions can be
found in \cite{Ellis}, where we can set $d=4$ again.

\section{Hadronic three jet cross sections}
\label{sec:3}

At HERA, positrons and protons are currently collided with energies of
$E_e = 27.5$ GeV and $E_p = 820$ GeV. The hadronic three jet cross section can
be calculated from the partonic cross section $\sigma_{12}^{345}$ with the help
of the factorization theorem,
\begin{eqnarray}
 \sigma_{ep}^{\rm 3-jet} (s_H) &=& \int \mbox{d}x_e f_e(x_e,M^2) \mbox{d}x_p
 f_p(x_p,M^2) \nonumber \\
 && \sigma_{12}^{345} (x_ex_ps_H,M^2).
\end{eqnarray}
$M$ is the factorization scale and $\sqrt{s_H} = \sqrt{(p_e+p_p)^2} = 300$
GeV is the hadronic center-of-mass energy. $f_e$ denotes the parton density in
the positron and is given by a convolution of the density of photons in the
positron with the density of partons in the photon,
\begin{equation}
 f_e(x_e,M^2) = \int \frac{\mbox{d}y}{y}f_e^\gamma(y)
 f_\gamma^{\gamma,q,g}(x_e/y,M^2).
\end{equation}
$f_e^\gamma(y)$ is given by the Weizs\"acker-Williams-Approximation \cite{WWA},
and we include the subleading terms calculated in \cite{Frixione}.
Experimentally the photon has a virtuality of $Q < 1$ GeV and carries a
fraction $0.2 < y < 0.8$ of the positron energy. For the parton densities in
the photon $f_\gamma^{\gamma,q,g}$ and proton $f_p^{q,g}$, we use the leading
order parametrizations by GRV \cite{GRV} and CTEQ4 \cite{CTEQ4} except where
stated otherwise. Asymptotically, the parton densities in the photon are of
${\cal O} (\alpha/\alpha_s)$, so when they are convoluted with the resolved
$2\rightarrow 3$ matrix elements of ${\cal O} (\alpha_s^3)$, the hadronic
cross section is of the same order as the direct contribution.
The strong coupling constant $\alpha_s(\mu)$ is calculated in
the one-loop approximation with five flavors and $\Lambda^{(5)} = 181$ MeV as
found in the global fit of the parton densities in the proton. The
renormalization and factorization scales are set to the maximum transverse jet
energy in the event, $\mu = M = \max(E_{T,1},E_{T,2},E_{T,3})$.

In the ZEUS three jet analysis, jets are defined by the KTCLUS algorithm in
inclusive mode \cite{KTCLUS}. This definition uses a parameter $R=1$ and
combines two hadronic clusters $i$ and $j$ if
\begin{equation}
  d_{ij} = \min(E_{T,i}^2,E_{T,j}^2)[(\eta_i-\eta_j)^2+(\phi_i-\phi_j)^2]/R^2
\end{equation}
is smaller than
\begin{equation}
  d_i = E_{T,i}^2.
\end{equation}
The transverse energy $E_T$, pseudorapidity $\eta$, and azimuthal angle $\phi$
of the combined cluster are calculated from the transverse energy weighted
sums of the two pre-clusters,
\begin{eqnarray}
 E_T  &=& E_{T,i}+E_{T,j}, \\
 \eta &=& [E_{T,i}\eta_i+E_{T,j}\eta_j]/E_T, \\
 \phi &=& [E_{T,i}\phi_i+E_{T,j}\phi_j]/E_T.
\end{eqnarray}
The longitudinal momentum fractions $x_p$ and $x_e$ of a parton in the
proton and electron can be determined from the observed three jet final state
through four momentum conservation
\begin{equation}
 x_{p,e} = \sum_{i=3}^{5} E_{T,i}e^{\pm\eta_i} / \sqrt{s_H},
\end{equation}
where the plus sign applies to the proton going into the positive $z$
direction and the jet pseudorapidities $\eta_i$ are defined in the
center-of-mass frame.
The longitudinal momentum fraction of a parton in the photon is
given by $x_{\gamma} = x_e/y$.
The three jets are required to have transverse energies of $E_{T,3},
E_{T,4} > 6$ GeV and $E_{T,5} > 5$ GeV, and their pseudorapidities must lie
within $|\eta_{3,4,5}| < 2.4$ in the laboratory frame \cite{Brussels}.
The KTCLUS algorithm is favored experimentally and in higher order
theoretical predictions since it contains no overlap or double counting
ambiguities. Leading order QCD predictions, however, lack the possibility to
implement or distinguish a particular jet definition, since each parton is
identified with a jet. 

For $N$ massless jets, one can choose $3N-4$ parameters that should span the
multijet parameter space, facilitate a simple interpretation within QCD, and
allow for a comparison of the $N-1$-jet to the $N$-jet cross section. In the
case of $N=3$, the conventional choices are the three jet mass $M_{\rm 3-jet}$,
required by ZEUS to be larger than 50 GeV, and four dimensionless parameters.
The Dalitz energy fractions
\begin{equation}
 x_i = \frac{2E_i}{M_{\rm 3-jet}}
\end{equation}
specify how the available energy is shared between the three jets. They are
ordered such that $x_3 > x_4 > x_5$. Since $x_3+x_4+x_5 = 2$, only $x_3$ and
$x_4$ are linearly independent. The third and fourth parameters are the
cosine of the scattering angle between the leading jet and the average beam
direction $\vec{p_{\rm AV}} = \vec{p_1}-\vec{p_2}$, where the incoming
parton 1 is the one with the highest energy in the laboratory frame,
\begin{equation}
 \cos\theta_3=\frac{\vec{p_{\rm AV}}\vec{p_3}}{|\vec{p_{\rm AV}}||\vec{p_3}|},
\end{equation}
and the angle between the three jet plane and the plane containing the
leading jet and the beam direction,
\begin{equation}
 \cos\psi_3=\frac{(\vec{p_3}\times\vec{p_{\rm AV}})(\vec{p_4}\times\vec{p_5})}
                 {|\vec{p_3}\times\vec{p_{\rm AV}}||\vec{p_4}\times\vec{p_5}|}.
\end{equation}
Instead of $\cos\psi_3$ we will investigate distributions in $\psi_3$ itself.
In the soft limit, where $E_5 \rightarrow 0$ and $x_{3,4} \rightarrow 1$,
$\cos\theta_3$ approaches the $2\rightarrow 2$ center-of-mass scattering angle
$\cos\theta^{\ast}$ thus relating three jet to two jet cross sections. 
$\cos\theta^{\ast}$ can be determined from the pseudorapidities of the two jets by
\begin{equation}
 \cos\theta^\ast = \tanh\left(\frac{\eta_3-\eta_4}{2}\right)
\end{equation}
and is related to the $2\rightarrow 2$ Mandelstam variables by
\begin{equation}
 t = -\frac{1}{2}s(1-\cos\theta^\ast)~~~\mbox{and}~~~
 u = -\frac{1}{2}s(1+\cos\theta^\ast).
\end{equation}
Of course, the third jet must not be too soft (or the hard jets not too hard)
to avoid soft singularities that would have to be absorbed into the
next-to-leading order dijet cross section. This is achieved by a cut on $x_3 <
0.95$. However, since the energy of a jet is always larger than its transverse
energy, the cut on $E_{T,5} > 5$ GeV already insures the absence of soft
singularities. The three jets also have to be well separated in phase space
from each other and from the incident beams to avoid initial and final state
collinear singularities. This is insured by a cut on $|\cos\theta_3| < 0.8$,
by the cuts in the pseudorapidities, and by the isolation cuts in the jet
definition.

\section{Numerical results}
\label{sec:4}

\begin{figure*}
 \begin{center}
  {\unitlength1cm
  \begin{picture}(12,17)
   \epsfig{file=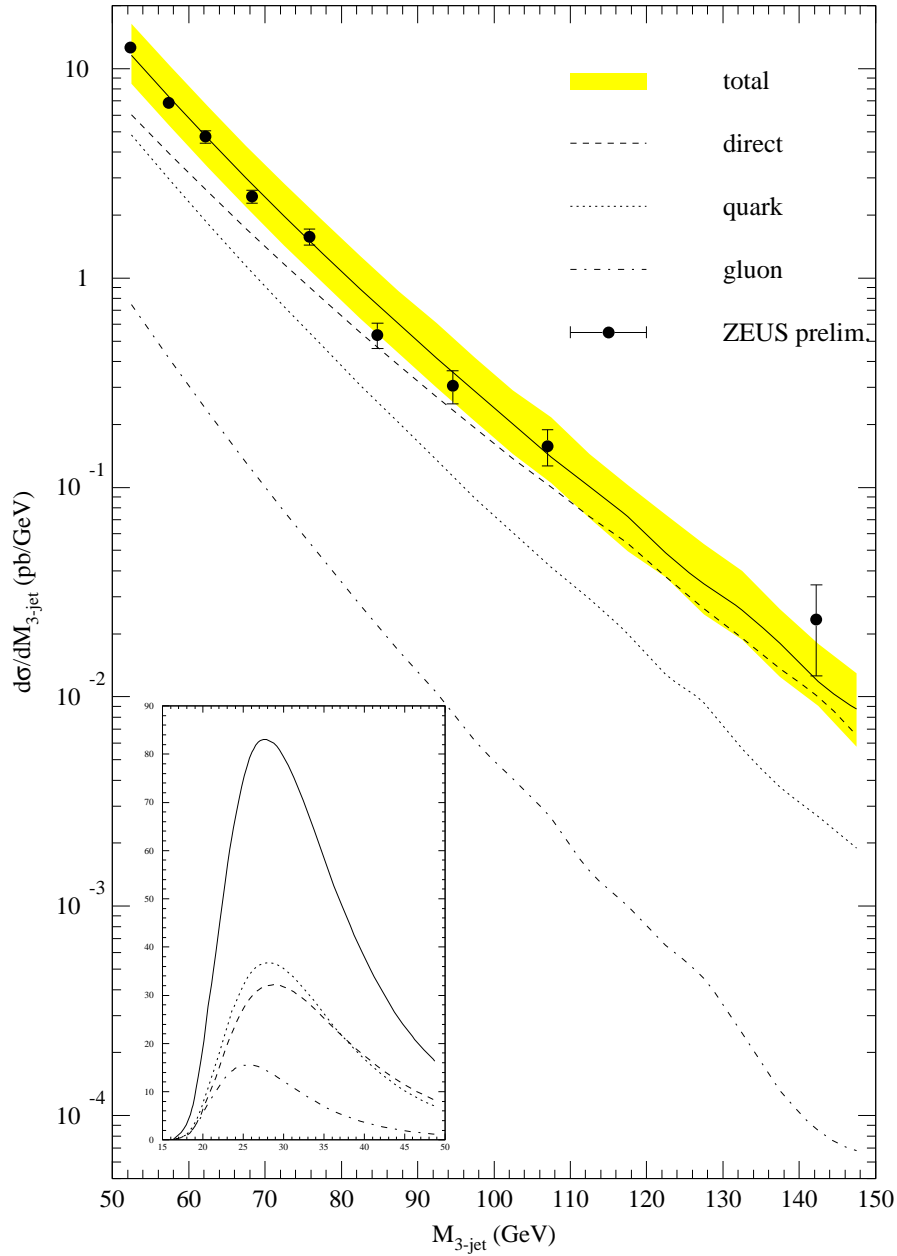,bbllx=60pt,bblly=95pt,bburx=495pt,bbury=715pt,%
           height=17cm,clip=}
  \end{picture}}
 \end{center}
\caption{Total cross section (full curve) for the photoproduction of three
         jets as a function of the three jet mass $M_{\rm 3-jet}$. We also
         show the variation
         of the absolute normalization due to the uncertainty in the
         scale choice (shaded band) and the contributions from direct
         photons (dashed), quarks (dotted), and gluons (dot-dashed) in the
         photon. The ZEUS data \cite{Brussels} agree well with the QCD
         prediction in shape and normalization within the statistical
         error.}
\label{fig:3}
\end{figure*}

Having explained the relevant theoretical details in the last two
sections, we can now turn to the presentation of numerical results for the
photoproduction of three jets at HERA. We first look for discrepancies
between theory and ZEUS data \cite{Brussels} in the three jet mass
distribution in Fig.~\ref{fig:3}.
We find that the total theoretical prediction (full curve) describes the
data well in shape and normalization. The agreement in normalization is,
however, to some degree coincidental since there is still an uncertainty
of a factor of two from the variation of the scales $\mu = M \in [0.5;2.0]
\times \max(E_{T,3}, E_{T,4},E_{T,5})$ (shaded band). This theoretical
uncertainty is much bigger than the statistical experimental
error of about 5\% at small $M_{\rm 3-jet}$ indicating
the need to implement next-to-leading order corrections in the three jet
cross section. Since these are not available yet, we will subsequently
resort to normalized distributions that are largely independent of both
higher order corrections and scale choices. In Fig.~\ref{fig:3} we also
show the individual contributions to the total cross section by direct photons
(dashed curve, contributes more than 50\%), by quarks (dotted curve, contributes
less than 40\%), and by gluons (dot-dashed curve, contributes less than 10\%) in
the photon. From the inset smaller figure it becomes clear that the quark-initiated
process gives the largest contribution for $M_{\rm 3-jet} < 36$ GeV. We also
computed but do not show here the individual contributions from quarks
and gluons in the proton and find the gluon to give the largest contribution
for $M_{\rm 3-jet}
< 80$ GeV. In the region of 36 GeV $< M_{\rm 3-jet} < 80$ GeV, the
photon-gluon fusion process accounts for one third of the total cross
section.

\begin{figure*}
 \begin{center}
  {\unitlength1cm
  \begin{picture}(12,8)
   \epsfig{file=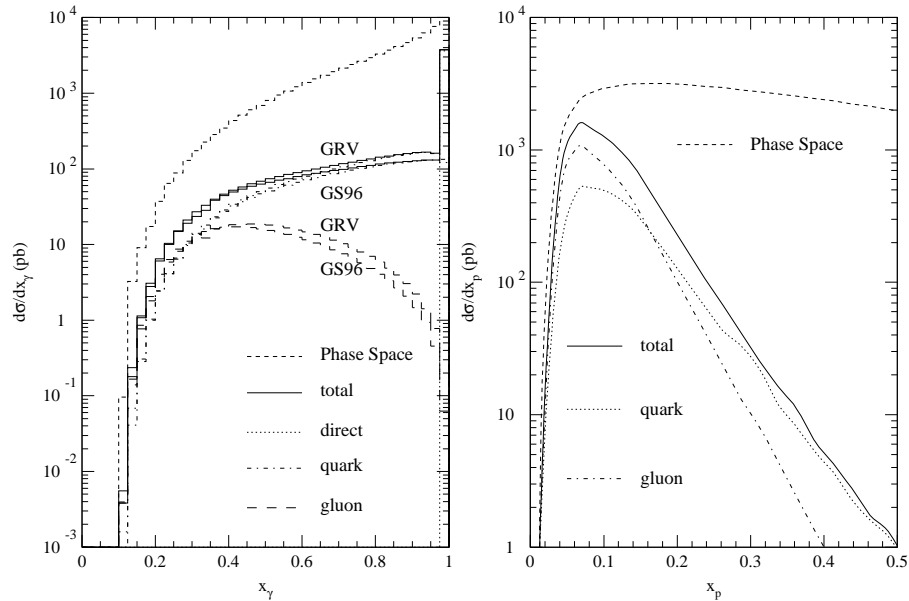,bbllx=520pt,bblly=85pt,bburx=100pt,bbury=715pt,%
           height=12cm,clip=,angle=270}
  \end{picture}}
 \end{center}
\caption{Total three jet cross sections per bin (full curves)
         as functions of the longitudinal momentum fraction $x_{\gamma}$
         of a direct photon (dotted), quark (dot-dashed), or gluon (long-dashed)
         in the photon (left) and of the longitudinal momentum fraction
         $x_p$ of a quark (dotted) or gluon (dot-dashed) in the proton (right).
         Also shown are the pure phase space distributions (dashed curves)
         which differ clearly from the dynamical QCD distributions.
}
\label{fig:4}
\end{figure*}

In Fig.~\ref{fig:4},
we show the total three jet cross section per bin as a function of
the longitudinal momentum fractions $x_{\gamma}$ of the partons in the photon
(left) and $x_p$ of the partons in the proton (right). These distributions
reflect the situation near the cut at $M_{\rm 3-jet} > 50$ GeV, since the three
jet cross section falls off exponentially with the three jet mass. In the left figure,
the total cross section (full curve) is presented together with the direct
peak at $x_{\gamma} = 1$ (dotted), which has been divided by the bin width,
and the quark (dot-dashed) and gluon (long-dashed) initiated processes. The
kinematic cuts restrict the phase space to $x_{\gamma} > 0.1$ (dashed). The
quark initial state is bigger than the gluon above $x_{\gamma} = 0.3$, and
both contributions are larger (smaller) in the GRV parametrization than in
GS96 \cite{GS96} when $x_{\gamma} > 0.35$ ($x_{\gamma} < 0.35$).
For the longitudinal momentum fraction $x_p$ of the parton in the proton
(right side of Fig.~\ref{fig:4}), the kinematic cuts restrict the phase space
to $x_p > 2.5 \cdot 10^{-2}$ and render contributions above $x_p > 0.5$
insignificant. The gluon (quark) in the proton dominates for $x_p < 0.16$
($x_p > 0.16$), and at low three jet masses, the gluon contributes about two
thirds of the total cross section.

\begin{figure*}
 \begin{center}
  {\unitlength1cm
  \begin{picture}(12,8)
   \epsfig{file=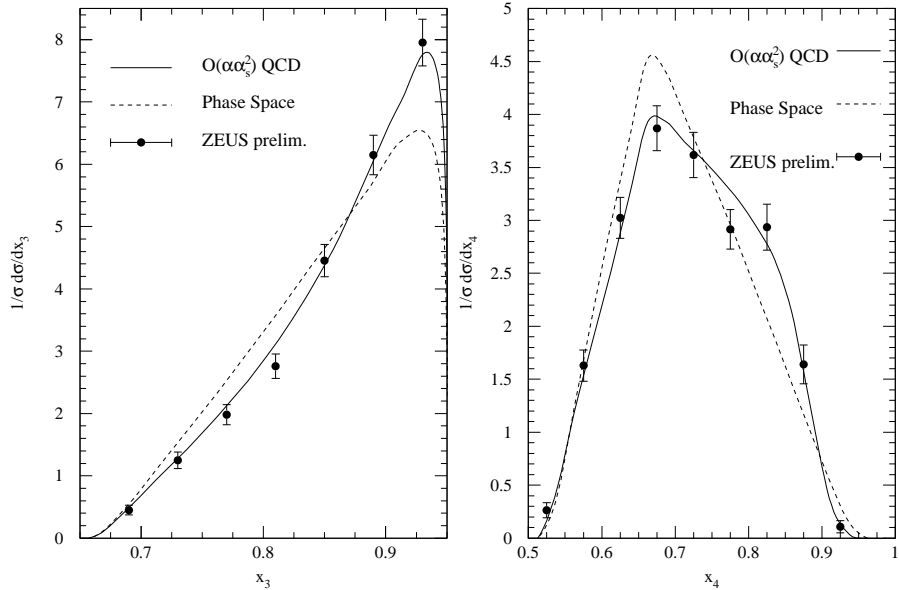,bbllx=520pt,bblly=85pt,bburx=100pt,bbury=715pt,%
           height=12cm,clip=,angle=270}
  \end{picture}}
 \end{center}
\caption{Dependence of the total cross section (full curves) and phase space
(dashed curves) on the energy fractions $x_3$ (left) and $x_4$ (right) of the
leading and next-to-leading jets. The ZEUS data \cite{Brussels} show a slight
preference of the full QCD prediction over phase space.}
\label{fig:5}
\end{figure*}

We now turn to normalized distributions in the four dimensionless three jet
parameters $x_3, x_4, \cos\theta_3,$ and $\psi_3$. Some of the theoretical
predictions have already been included in the presentation of the experimental
data \cite{Brussels}. Fig.~\ref{fig:5}
shows the three jet cross section as a function of
the energy fractions of the leading (left) and next-to-leading (right) jet
$x_3$ and $x_4$, normalized to the total cross section. The prediction from
the ${\cal O}(\alpha\alpha_s^2)$ QCD matrix elements (full curves) are rather
similar to the pure phase space distributions (dashed curves), but the data
slightly prefer the QCD predictions. The statistical error bars are still too
large for a definite conclusion, however. We have also studied the relative
contributions from the different partonic sub-channels. They are flat in
$x_3$, but photon-quark scattering clearly dominates when $x_4 \simeq x_3
\rightarrow 1$.

\begin{figure*}
 \begin{center}
  {\unitlength1cm
  \begin{picture}(12,17)
   \epsfig{file=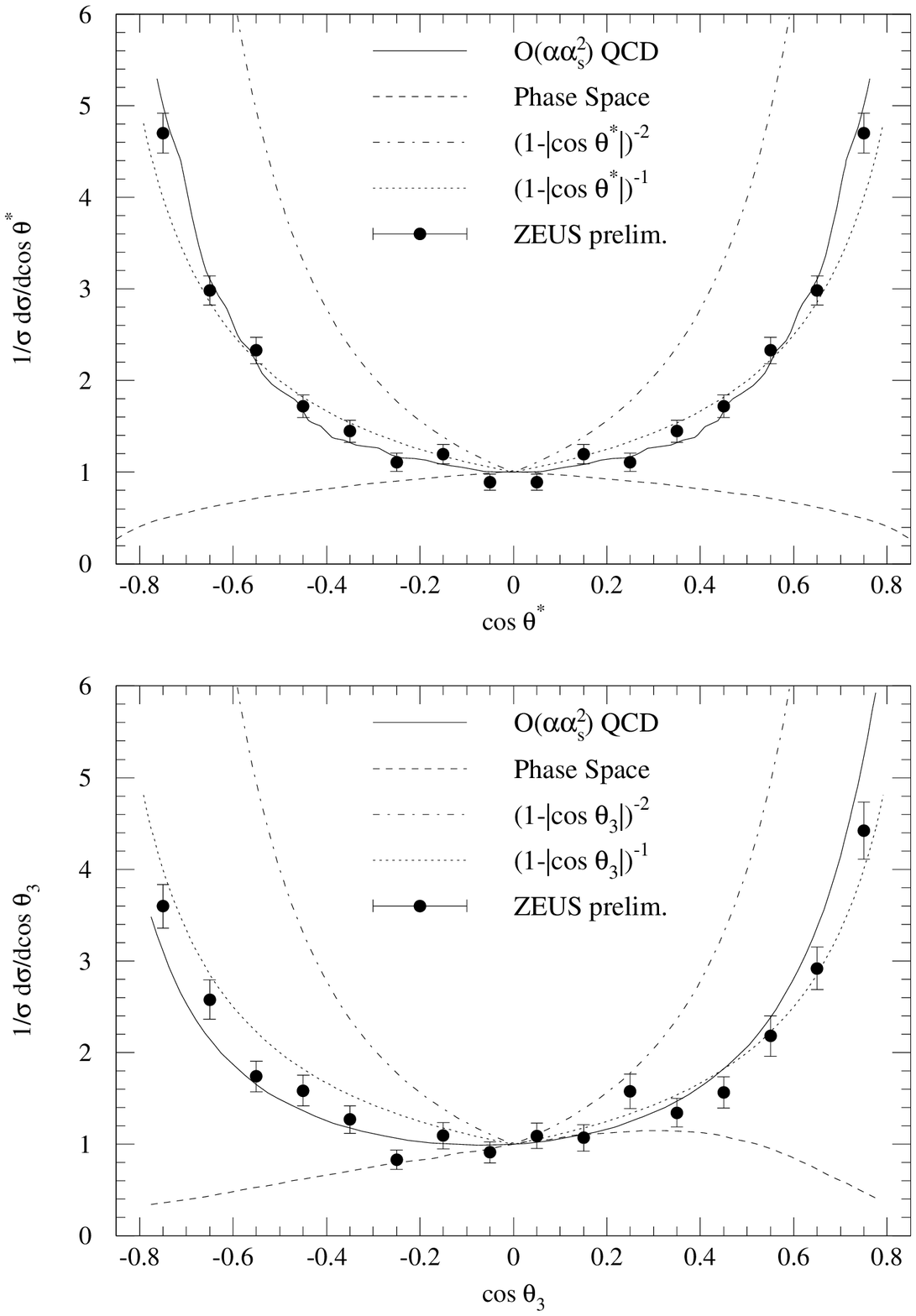,bbllx=60pt,bblly=95pt,bburx=495pt,bbury=715pt,%
           height=17cm,clip=}
  \end{picture}}
 \end{center}
\caption{Cross sections for the photoproduction of two (upper plot) and three
(lower plot) jets as functions of the center-of-mass and fastest jet scattering
angles normalized at $\cos\theta=0$.
The ZEUS dijet \cite{ZEUSDijet} and three jet data \cite{Brussels} are
well described by the QCD predictions (full curves), but not by the pure phase
space distributions (dashed curves), and they favor the single $1/t$ pole for
massless fermion exchange (dotted curves) over the double $1/t^2$ pole for
massless boson exchange (dot-dashed curves) in the $t$-channel.
}
\label{fig:6}
\end{figure*}

The distribution in the cosine of the scattering angle of the fastest jet 
$\cos\theta_{3}$ is displayed in the lower plot of Fig.~\ref{fig:6}
together with the distribution in the dijet scattering angle $\cos\theta^\ast$
(upper plot) measured earlier by ZEUS \cite{ZEUSDijet} and compared to theoretical
predictions in \cite{Klasen}. They are both of ${\cal O}
(\alpha\alpha_s^2)$ which is the leading order for the three jet case but
next-to-leading order for the dijet case. The wiggles in the latter stem from
limited statistics in the Monte Carlo integration. Of course, the cuts in the
dijet and three jet cross sections differ slightly, but the most important cuts
on the jet mass $M_{\rm 2-jet} > 47$ GeV and the scattering angle
$|\cos\theta^{\ast}| < 0.8$ are in fact very similar. The differences will also
mainly affect the normalization and should be insignificant in distributions
normalized at $\cos\theta = 0$ as they are presented here. Let us first
concentrate on the dijet case. As expected it is symmetrical in $\cos\theta^{\ast}$.
The data show good agreement with the QCD prediction (full curve) but clearly
disagree with the pure phase space distribution (dashed curve). It might be
surprising at first sight that the data also disagree with the Rutherford
scattering form at small angle, $(1-|\cos\theta^{\ast}|)^{-2}$. On the other hand,
the data agree very well with the less singular form $(1-|\cos\theta^{\ast}|)^{-1}$.
This can be understood if we remind ourselves that Rutherford scattering is
characteristic of the exchange of a massless vector boson in the $t$-channel
\begin{equation}
 \overline{|{\cal M}|^2} \propto t^{-2} = \left[ -\frac{1}{2} s (1 -
 \cos\theta^\ast) \right]^{-2}
\end{equation}
as present in most of the resolved (parton-parton) scattering processes
whereas both direct processes proceed through a massless {\em fermion}
exchange in the $t$-channel with less singular behavior,
\begin{equation}
 \overline{|{\cal M}|^2} \propto t^{-1} = 
   \left[ -\frac{1}{2} s(1-\cos\theta^\ast)\right]^{-1}.
\end{equation}
In addition, the process $\gamma q \rightarrow qg$ has an $s$-channel contribution
without any singular behavior. Therefore, the $\cos\theta^{\ast}$ distribution
provides evidence that direct processes and photon-gluon fusion in particular
are the most important subprocesses in this kinematic region. Turning now to the
three jet cross section,
we observe that both the ${\cal O}(\alpha\alpha_s^2)$
QCD (full curves) and phase space distributions (dashed curves) are asymmetric
in $\cos\theta_3$ which motivates why we kept the sign in the abscissa and did not
present curves in the absolute value $|\cos\theta|$ as one usually does in the dijet
case. Again the data clearly prefer QCD (full) over phase space (dashed)
and the single pole in $t$
(dotted) over the double pole (dot-dashed). In fact, the dependence of the three jet
cross section on the scattering angle of the fastest jet is strikingly similar to
the dijet distribution in the center-of-mass scattering angle. The deviation in the
backward region of the three jet case $\cos\theta_3 < 0$, when the pseudorapidity of
the
fastest jet $\eta_3 < \eta_{4,5}$ is tending towards the direction of
the photon beam, can be traced to a
large extent to the asymmetric phase space. Since three jet production proceeds
dominantly through single gluon bremsstrahlung, we can relate this result to the
underlying $2\rightarrow 2$ process with the interpretation given there. In fact
it is interesting to note that the CDF collaboration compared their first three
jet data directly to two jet
predictions when three jet predictions were still unavailable \cite{CDFDijet}.

\begin{figure*}
 \begin{center}
  {\unitlength1cm
  \begin{picture}(12,8)
   \epsfig{file=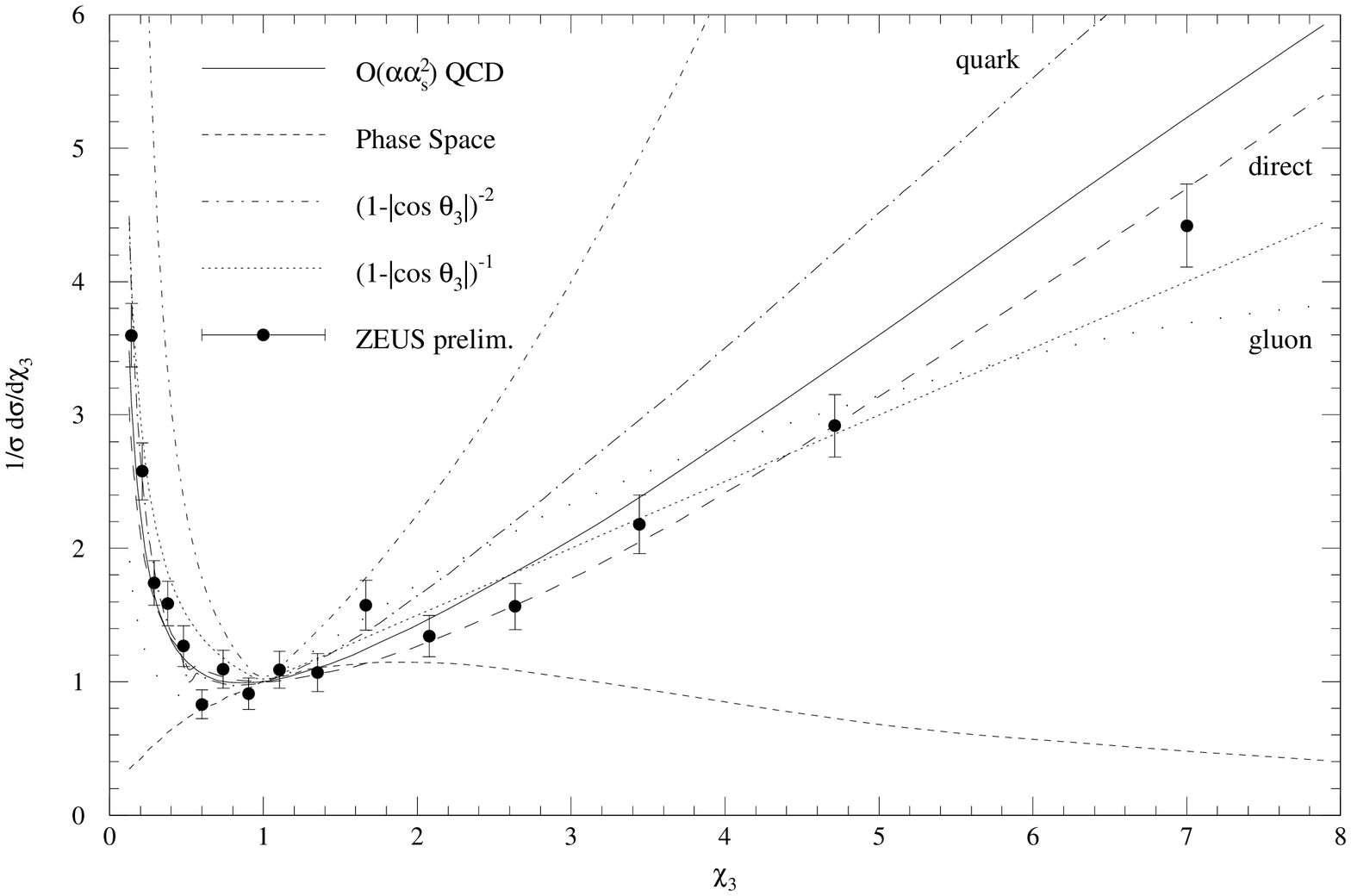,bbllx=520pt,bblly=85pt,bburx=100pt,bbury=715pt,%
           height=12cm,clip=,angle=270}
  \end{picture}}
 \end{center}
\caption{Dependence of the total three jet cross section
on the angular variable $\chi_3$. The ZEUS data \cite{Brussels}
agree with the QCD prediction
(full curve), which is dominated by the direct contribution (long-dashed), and with
the single pole in $t$ (dotted). They exclude the pure phase space (dashed), quark
(long-dash-dotted) and gluon (wide-dotted) initiated distributions as well as the
double pole in $t$ (dot-dashed).}
\label{fig:7}
\end{figure*}

To study deviations from the
$(1-\cos\theta)^{-2}$ behavior it is convenient to plot the
data in terms of the variable \cite{Combridge}
\begin{equation}
 \chi = \frac{1+\cos\theta}{1-\cos\theta}
\end{equation}
which is related to the experimentally measured pseudorapidities by
\begin{equation}
 \chi = e^{\eta_1-\eta_2}
\end{equation}
and removes the Rutherford singularity such that
\begin{equation}
 \frac{\mbox{d}\sigma}{\mbox{d}\chi} \sim \mbox{constant}
\end{equation}
if one ignores the angular dependence and scaling violation in the coupling constant and the
parton densities. In Fig.~\ref{fig:7}
we present distributions in the angular variable $\chi$ of the fastest of three jets,
normalized at $\chi = 1$ (i.e. $\cos\theta = 0$). The data show again a clear
distinction from both pure phase space and Rutherford scattering
at small angle $(1-|\cos\theta|)^{-2}$ but agree very well with
${\cal O}(\alpha\alpha_s^2)$ QCD prediction and also with the form
$(1-|\cos\theta|)^{-1}$. The QCD prediction includes, of course, scaling
violation both in the strong coupling and in the parton densities. Absence
of scaling violations would result in a constant prediction and would clearly
disagree with the data. As before, the full QCD prediction is
dominated by direct photon scattering. Therefore the long-dashed curve lies close
to the full curve, whereas the processes initiated by the quark in the photon
(long-dash-dotted curve) give a steeper shape as expected. The gluon initiated
processes (wide-dotted curve) have an even different shape. The total resolved
contribution (not shown), i.e. the sum of quark and gluon initiated contributions,
lies between the quark initiated and total curves and accounts for the remaining
difference of the direct initiated and total cross section.

\begin{figure*}
 \begin{center}
  {\unitlength1cm
  \begin{picture}(12,8)
   \epsfig{file=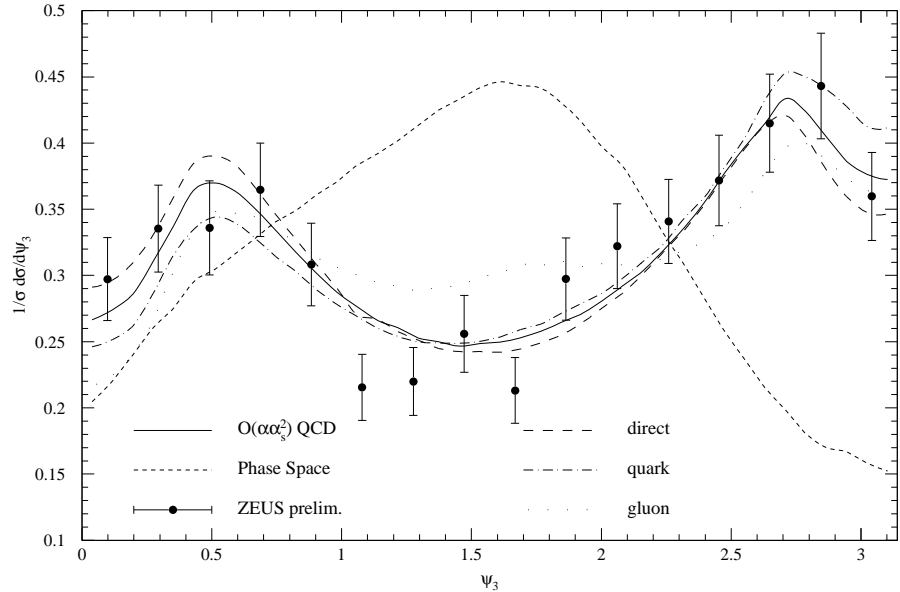,bbllx=520pt,bblly=85pt,bburx=100pt,bbury=715pt,%
           height=12cm,clip=,angle=270}
  \end{picture}}
 \end{center}
\caption{Dependence of the total three jet cross section (full curve), the
pure phase space (dashed curve), and the individual contributions from the
partons in the photon (long-dashed, long-dash-dotted, and wide-dotted curves)
on the angle between the three jet plane and the plane containing the
leading jet and the average beam direction. The ZEUS data \cite{Brussels}
rule out the pure phase space and gluon initiated distributions.}
\label{fig:8}
\end{figure*}

Finally we investigate the dependence of the three jet cross section normalized
to the total cross section on the angle between the three jet plane and the
plane containing the leading jet and the average beam direction in Fig.~\ref{fig:8}.
The full QCD curve again agrees well with the data as do the contributions from direct
photons (long-dashed curve) and quarks in the photon (long-dash-dotted curve), whereas the pure
phase space (dashed curve) has a completely different shape. The gluon in the photon
(wide-dotted curve) gives a slightly different shape.

\section{Summary}
\label{sec:5}
In this paper we have presented three jet cross sections in photoproduction using
exact leading order matrix elements. The normalization and shape of the QCD predictions
show good agreement with ZEUS data, although the normalization is subject to rather
large scale uncertainties. None of the distributions can be explained by the
three jet phase space alone. Of the different direct and resolved contributions,
direct and in particular photon-gluon fusion processes are the most important
subprocesses. The fastest-jet
scattering angle distribution looks very similar to the dijet center-of-mass
scattering angle distribution indicating that three jet production proceeds
mainly through single bremsstrahlung. The data are not consistent with
Rutherford scattering for the exchange of a massless boson but with the
less singular form for the exchange of a massless fermion in the $t$-channel.

\begin{acknowledgement}
Work in the High Energy Physics Division at Argonne National Laboratory
is supported by the U.S. Department of Energy, Division of High Energy
Physics, Contract W-31-109-ENG-38. The author thanks Laurel Sinclair
and Esther Strickland for making the preliminary ZEUS data available
to him and Gustav Kramer for useful comments on the manuscript.
\end{acknowledgement}

\end{document}